\documentclass[prl,twocolumn,amsmath,superscriptaddress,longbibliography]{revtex4-1}
\usepackage{graphicx,subfigure}
\usepackage{bm}
\newcommand{\bracket}[1]{\langle#1\rangle}
\newcommand{\ket}[1]{|#1\rangle}

\DeclareMathOperator{\IM}{Im}
\DeclareMathOperator{\TR}{Tr}

\usepackage[colorlinks=true, letterpaper=true, pdfstartview=FitV,
  linkcolor=blue, citecolor=blue, urlcolor=blue]{hyperref}

\begin{document}

\title{Topological Inverse Faraday Effect in Weyl Semimetals}

\author{Yang Gao}

\affiliation{Department of Physics, University of Science and Technology of China,
  Hefei, Anhui 100049, China}

\author{Chong Wang}

\affiliation{Department of Physics, Carnegie Mellon University,
  Pittsburgh, PA 15213, USA}
  
\author{Di Xiao}

\affiliation{Department of Physics, Carnegie Mellon University,
  Pittsburgh, PA 15213, USA}

\date{\today}

\begin{abstract}
We demonstrate that in Weyl semimetals, the momentum-space helical spin texture can couple to the chirality of the Weyl node to generate a frequency-independent magnetization in response to circularly polarized light through the inverse Faraday effect.  This frequency-independence is rooted in the topology of the Weyl node. Since the helicity and the chirality are always locked for Weyl nodes, this effect is not subject to any symmetry constraint. Finally, we show that the photoinduced frequency-independent magnetization is robust against lattice effect and has a magnitude large enough to realize ultrafast all-optical magnetization switching below picosecond.
\end{abstract}

\maketitle

Weyl semimetals, which host topologically protected Weyl nodes with linear dispersion, have been the focus of condensed matter physics in recent years~\cite{Armitage2018}. One of the most celebrated features of Weyl semimetals is that each Weyl node acts as a magnetic monopole in the momentum space, emitting the Berry curvature as the effective magnetic field, and possessing a quantized chirality, defined as the monopole charge. This chiral feature leads to many intriguing experimental consequences~\cite{Wan2011,Son2013,Zyuzin2012,Lv2015,Xu2015,Chan2016,Juan2017,Armitage2018}. In addition to its chirality, since Weyl semimetals often have sizable spin-orbital coupling, each Weyl node can also possess a helical spin texture in the momentum space. However, the implication of such helical spin texture is less explored theoretically and experimentally.

In this Letter, we study the role of the helical spin texture in the inverse Faraday effect~(IFE) in Weyl semimetals. The IFE refers to a photoinduced static magnetization that switches sign with the light chirality.  It has been studied previously in general context~\cite{Pershan1966,Vahaplar2009,Battiato2014,Freimuth2016,Berritta2016}, but its distinct features in Weyl semimetals, especially its connection to the topology of the Weyl node, have not been thoroughly explored.
Here we show that the IFE in Weyl semimetals results from the combined effects of the chirality and helicity of the Weyl node~(see Eq.~\eqref{eq_trbeta}): the chirality flux density determines the coupling to chiral light whereas the helicity flux density determines the angular momentum transfer. We show that the helicity, similar to the chirality, is an intrinsic property of the Weyl node owing to its topological nature.  Remarkably, the photoinduced magnetization is frequency-independent. We further evaluate the IFE in a lattice model of Weyl semimetals and show that the insensitivity to the light frequency is robust against lattice effect.

In general, the IFE provides an efficient mechanism to control the material magnetization without external magnetic fields~\cite{Pershan1966,Vahaplar2009,Battiato2014,Freimuth2016,Berritta2016}.  Realizing the above frequency-independent IFE in Weyl semimetals hence represents a highly desirable advancement as it eliminates the need to fine tune the light frequency close to resonance.  This frequency-independent IEF is also very robust as it can exist in both non-centrosymmetric and magnetic Weyl semimetals. For magnetic Weyl semimetals, in a typical experimental setup~\cite{Stanciu2007,Huisman2016}, we estimate that the photoinduced magnetization can flip the magnetization on the order of 120~fs with a 50~fs laser pulse of fluence 1 ${\rm mJ/cm^2}$.  This shows that it is feasible to utilize such photoinduced frequency-independent magnetization in Weyl semimetals to realize ultrafast all-optical magnetization switching, highlighting the potential of Weyl semimetals in opto-spintronic applications.

\textit{The inverse Faraday effect.}---Under the illumination of circularly polarized light, crystals are expected to exhibit a photo-induced magnetization, whose time-independent part $\bm M$ is dominated by the second order optical response~(i.e. the IFE), \begin{equation}\label{eq_pm}
M_i=\beta_{ij}(\omega)[i\bm E(\omega)\times \bm E^\star(\omega)]_j \;.
\end{equation}
For right/left-circularly polarized light propagating along the $j$-th direction, $[i\bm E(\omega)\times \bm E^\star (\omega)]_j=\pm |\bm E(\omega)|^2$. 

Different components of $\beta_{ij}$ have different symmetry requirements. Since both $\bm M$ and $\bm E\times \bm E^\star$ transform as axial vectors, one immediately realizes that the off-diagonal part of $\beta_{ij}$ requires the breaking of two mirror symmetries, with the corresponding mirror planes normal to $\bm M$ and the light propagation direction, respectively; but it is insensitive to time reversal or spatial inversion operations. 
In contrast, the diagonal part of $\beta_{ij}$ does not need to break any mirror, spatial inversion, or time reversal symmetry. Specifically, the trace $\TR\beta_{ij}$ simply transforms as a scalar under point group operations. 

The focus of this work is the diagonal part of $\beta_{ij}$. Without loss of generality, let us consider $\beta_{zz}$. From the above symmetry analysis, we see that $\beta_{zz}$ represents a robust effect not subject to any symmetry constraints, and can exist in a wide variety of materials. The expression of $\beta_{zz}$ can be derived using standard second-order response theory~\cite{Sipe2000,Morimoto2016,Gao2020}, with the result reading~(we set $e=\hbar=\mu_B=1$ hereafter for simplicity)
\begin{equation}\label{eq_betaf}
\begin{split}
\beta_{zz}=&\frac{2i}{\omega^2}\sum_{\ell,m,n}\int \frac{d\bm k}{8\pi^3} \frac{(v_y)_{m\ell}(v_x)_{\ell n}-(x\leftrightarrow y)}{\omega_{nm}+i/\tau_0} \\
&\times (G_{\ell n}+G_{m\ell})(s_z)_{nm} \;,
\end{split}
\end{equation}
where $(v_i)_{m\ell}$ and $(s_z)_{nm}$ are the velocity and spin matrix element in the band basis, respectively, $\omega_{nm}=\varepsilon_n-\varepsilon_m$, and $G_{\ell n}=(f_\ell -f_n)/(\omega_{\ell n}-\omega -i/\tau_0)-(\omega\rightarrow -\omega)$ with $f_n$ being the Fermi-Dirac distribution function for band $n$. 

Similar to the circular photogalvanic effect~\cite{Sipe2000,Juan2017}, the dominant contribution to $\beta_{zz}$ comes from terms with $n=m$, as generally $\omega_{nm} \gg 1/\tau_0$ for $n\neq m$. Keeping only $n = m$ terms, Eq.~\eqref{eq_betaf} can be reduced to~\cite{suppl}
\begin{align}\label{eq_betaz}
\beta_{zz}&=- \tau_0 \sum_{\ell, n} \int \frac{d\bm k}{2\pi^2} (\Omega_z)_{n \ell} (\Delta s_z)_{n\ell}\delta(\omega_{\ell n}-\omega) \;,
\end{align}
where $(\Omega_z)_{n\ell}=-2 \IM [\bracket{u_n|i\partial_{k_x}|u_\ell} 
\bracket{ u_\ell|i\partial_{k_y}|u_n}]$ and $(\Delta s_z)_{n\ell}=\bracket{u_n|s_z|u_n}-\bracket{u_\ell|s_z|u_\ell}$. For a two-band model, $(\Omega_z)_{n\ell}$ reduces to the Berry curvature in band $n$. If we replace the spin difference $\Delta \bm s$ by the velocity difference, we recover the response coefficient for the circular photogalvanic effect~\cite{Sipe2000,Juan2017}.

\textit{IFE in Weyl semimetals}.---We now evaluate the response function in Eq.~\eqref{eq_betaz} for an isotropic Weyl node, described by the Hamiltonian
\begin{align}\label{eq_ham1}
\hat{H}=\chi v \bm k\cdot \bm \sigma \;,
\end{align}
where  $\chi=\pm 1$ is the chirality of the Weyl node, $v$ is the Fermi velocity, and $\bm \sigma$ operates in the pseudospin space. The role of the chirality and the helicity can be most easily seen in the trace of $\beta_{ij}$, which, for the isotropic Weyl Hamiltonian, reads~\cite{suppl}
\begin{align}\label{eq_decomp}
{\rm Tr}\beta_{ij}=-\frac{\tau_0}{v^2}  \int \frac{d\bm k}{2\pi^2} (\bm \Omega\cdot \bm v)(\bm v\cdot \Delta \bm s)\delta(2\varepsilon-\omega)\;,
\end{align} 
where $\bm \Omega=\chi v^3 \bm k/(2\varepsilon^3)$ is the Berry curvature, $\bm v=v^2 \bm k/\varepsilon$ is the velocity in the conduction band with $\varepsilon =v |\bm k|$ being the band energy, $\Delta \bm s=\bracket{u_c|\bm s|u_c} - \bracket{u_v|\bm s|u_v}$ is the $\bm k$-resolved spin difference between the conduction and the valence band.

\begin{figure}[t]
  \begin{minipage}{\linewidth}
  \centering
  \includegraphics[width=\columnwidth]{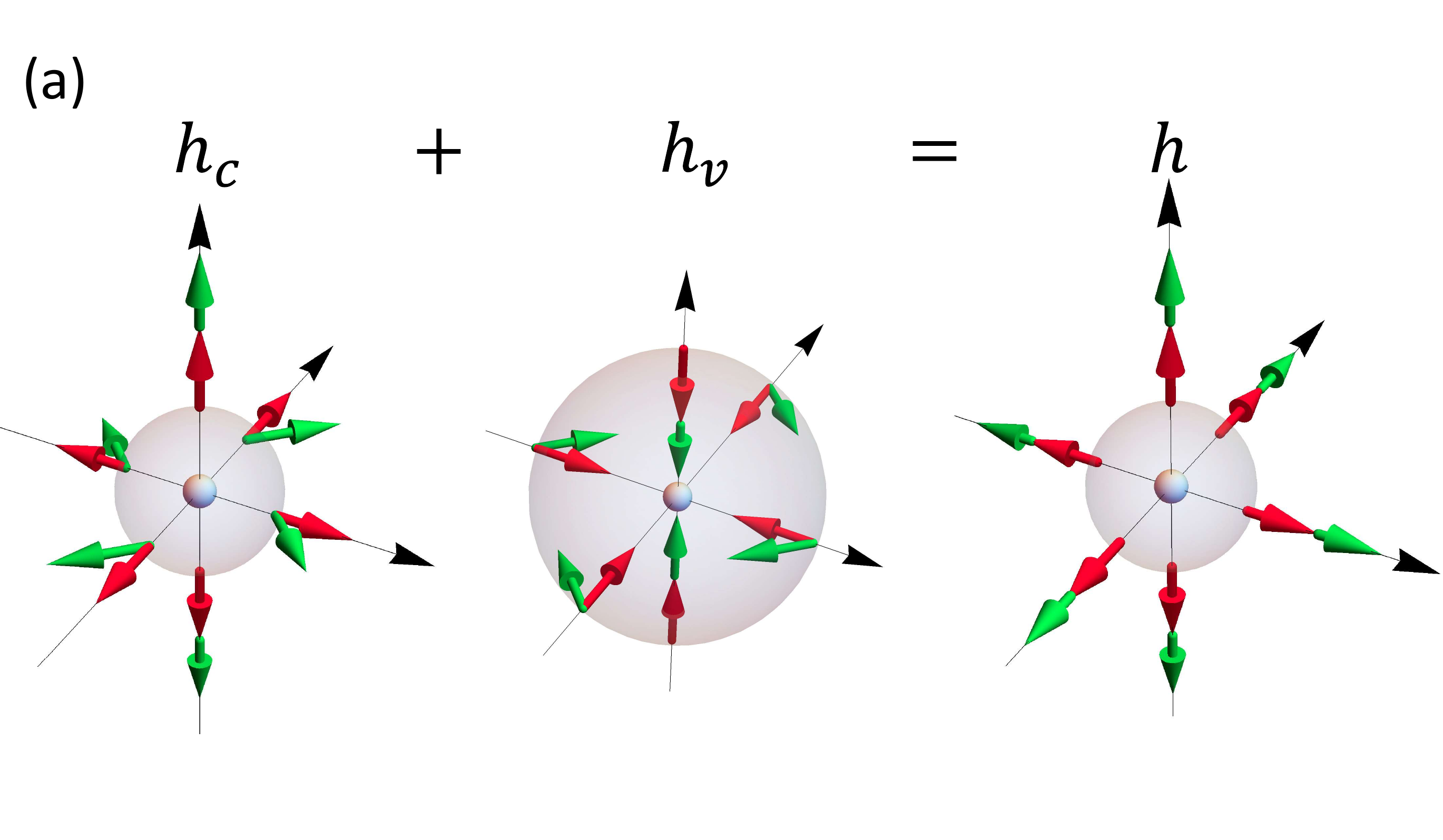}
  \end{minipage}
  \quad\\
  \begin{minipage}{\linewidth}
  \centering
  \includegraphics[width=\columnwidth]{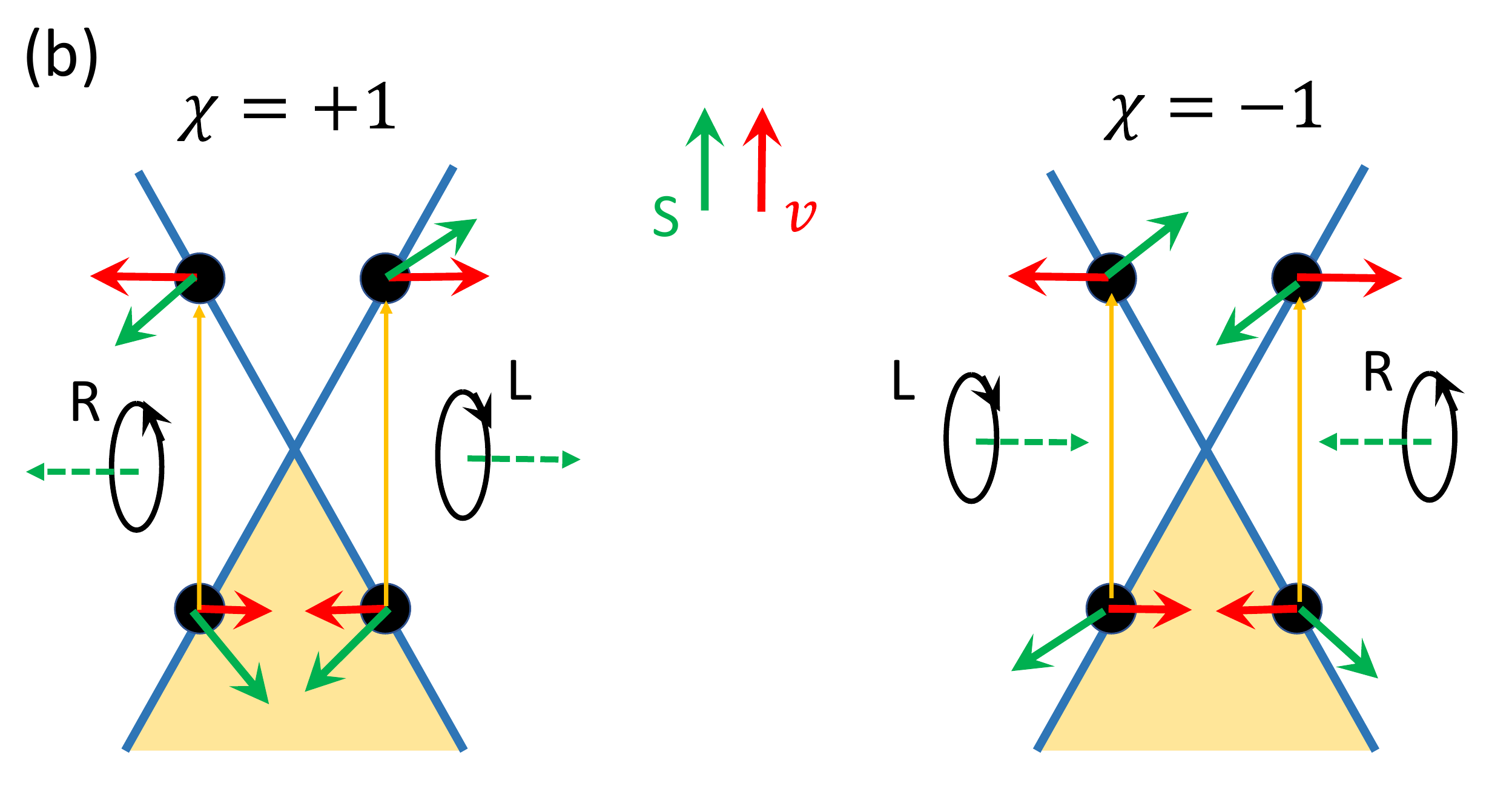}\\
  \end{minipage}
  \caption{Momentum-space helical spin texture~(a) and the schematic diagram of the inverse Faraday effect for a pair of Weyl nodes with opposite chiralities~(b). For (a), the three plots shows the spin texture for helicity in the conduction~($h_c$) and valence band~($h_v$), and their sum~($h$). In both (a) and (b), the red and green arrow stand for the band velocity and spin direction. In (b), the coupling to the chiral lights are determined by the Berry curvature in the conduciton band, which is parallel~(for $\chi=+1$) and antiparallel~(for $\chi=-1$) to the velocity. The dashed green arrow in (b) shows the angular momentum of chiral lights, which is transferred to the Weyl Fermions, accounting for the difference between the green arrows in the conduction and valence band. Therefore, from (b) one finds that the chirality and helicity are always locked, collaborating constructively to yield a nontrivial induced magnetization from chiral lights. }\label{fig_fig0}
\end{figure}

From Eq.~\eqref{eq_decomp}, one immediately finds that the IFE is determined by two factors, i.e., $\bm \Omega \cdot \bm v$ and $\bm v\cdot \Delta \bm s$. The first factor $\bm \Omega \cdot \bm v$ can be interpreted as the flux density of the chirality as the latter is given by
\begin{align}\label{eq_chi}
\chi=\frac{1}{2\pi}\int \bm \Omega\cdot \bm v \delta(\varepsilon-\varepsilon_0) d\bm k\,,
\end{align}
where the integration is over a constant energy surface in the conduction band.
It is well known that the Berry curvature couples to the light chirality by generating the difference in the oscillator strength for left/right circularly polarized light~\cite{Souza2008}, hence deciding the favorite light chirality in the optical transition. This is illustrated in Fig.~\ref{fig_fig0}(b), where the flipping of the Berry curvature direction implies coupling to opposite light chirality.

Interestingly, the second factor $\bm v\cdot \Delta \bm s$ can be identified as the flux density of the helicity. The helicity is originally defined as the projection of spin on the momentum $\bm k$. For the Weyl node in Eq.~\eqref{eq_ham1}, $\bm v= v\bm k/|\bm k|$ is parallel to $\bm k$; we can thus define the helicity $h$ as the projection of spin on $\bm v$. 
More precisely, we define
\begin{equation}\label{eq_hel}
h = \frac{v^2}{4\pi\varepsilon_0^2}\int \bm v\cdot \Delta \bm s \delta(\varepsilon-\varepsilon_0)  d\bm k\,,
\end{equation}
where the prefactor $v^2/(4\pi\varepsilon_0^2 )$ is chosen so that: (1) it makes $h$ dimensionless; and (2) it has the form $1/[vg(\varepsilon_0)]$ with $g(\varepsilon_0)=\int \delta(\varepsilon-\varepsilon_0) d\bm k$ being the density of states at energy $\varepsilon_0$, thus ensuring $|h|\le |\Delta \bm s|$. 

A nonzero helicity indicates a nontrivial spin texture. To appreciate this, we define the conduction and valence band helicity $h_\alpha$ by replacing $\Delta\bm s$ in Eq.~\eqref{eq_hel} with $\langle u_\alpha|\bm s|u_\alpha\rangle$, where $\alpha = c,v$ for the conduction and valence band, respectively.  Then $h$ can be written as $h=h_c+h_v$. An example of the spin texture for $h_c$, $h_v$, and the resulting $h$ is shown in Fig.~\ref{fig_fig0}(a). 

From the viewpoint of angular momentum conservation, the photoinduced magnetization requires the transfer of the angular momentum from chiral lights to Weyl Fermions~\cite{Woodford2009}. The degree of such angular momentum transfer is quantified by the helicity $h$. Since $\Delta \bm s$ is the spin difference between the conduction and the valence band, it is obvious that when the conduction and valence band are in the same spin state, i.e., $\Delta \bm s=0$, the angular momentum of light cannot be transferred, and hence the IFE is forbidden. Figure.~\ref{fig_fig0} (a) and (b) exemplify the most efficient angular momentum transfer, in which the spin difference is always parallel to the band velocity.

The chirality and the helicity together determine how the IFE behaves under symmetry operations. Since both $\Delta \bm s$ and $\bm \Omega$ transform as an axial vector, one immediately finds that by going from one Weyl node to its conjugate partner, both chirality and helicity flip sign. In other words, chirality and helicity are always locked for Weyl nodes. As a result, the photoinduced magnetizations from  Weyl nodes with opposite chiralities add up, as illustrated in Fig.~\ref{fig_fig0}(b).

We now proceed to calculate the IFE. 
Let us label the basis in the pseudospin space by $|+\rangle$ and $|-\rangle$. The helicity reads~\cite{suppl}
\begin{equation}\label{eq_fhel}
h = \frac{\chi}{3}[\bracket{+|s_z|+} - \bracket{-|s_z|-}
+ (\bracket{+|s_+|-} + \text{c.c.})]\;,
\end{equation}
where $s_+=s_x+is_y$. Therefore, the helical spin texture is eventually translated to the spin matrix elements in the basis of the Weyl Hamiltonian. We find that the helicity $h$ is independent of the energy $\epsilon_0$, which is inherited from the topology of the Weyl node, as one can show that $\bm v\cdot \Delta \bm s\propto \bm v\cdot \bm \Omega$~\cite{suppl}. The helicity can even be quantized in a Kramers-Weyl node with $\bm \sigma$ in Eq.~\eqref{eq_ham1} describing the real spin~\cite{Chang2018}; in this case $h=\chi$. 

This energy-independence of $h$ highlights the helicity as an intrinsic property of the Weyl node. Just like a real Fermion carries electric charge and spin, the Weyl Fermion can carry a chiral charge and a helicity, whose product generates the IFE. To be specific, when the Fermi energy is at the Weyl point, ${\rm Tr}\beta_{ij}$ for the isotropic Weyl node reads ~\cite{suppl},
\begin{align}\label{eq_trbeta}
{\rm Tr}\beta_{ij}&=-\frac{\tau_0}{2\pi v}\chi h\;.
\end{align}
This is a key result of this work. One immediately recognizes that ${\rm Tr}\beta_{ij}$ is independent of the light frequency, owing to the topological implication in $\chi$ and $h$. The magnitude of ${\rm Tr}\beta_{ij}$ can be tuned in principle, through varying $\tau_0$, $v$, or the spin structure in $h$. If the Fermi energy $\mu$ is away from the Weyl point, optical transition is forbidden for light frequency lower than $2|\mu|$.  The topological IFE in Eq.~\eqref{eq_trbeta} is recovered once the frequency is above $2|\mu|$.

In reality, the Weyl node may be anisotropic.  For example, the Fermi velocity may be different in the three spatial directions, corresponding to a model Hamiltonian $\hat{H}_a=(v_x\sigma_x k_x+v_y\sigma_y k_y+v_z\sigma_z k_z)\chi$.  In this case, one can show that the frequency-independence still persists, with the diagonal part of $\beta_{ij}$ given by~\cite{suppl}
\begin{equation} \label{eq_totbeta}
\begin{split}
\beta_{xx}&=-\frac{\tau_0}{6\pi v_x}(\bracket{+|s_x|-} + \text{c.c.}) \;, \\
\beta_{yy}&=-\frac{\tau_0}{6\pi v_y}(\bracket{+|is_y|-} + \text{c.c.}) \;, \\
\beta_{zz}&=-\frac{\tau_0}{6\pi v_z}(\bracket{+|s_z|+} 
- \bracket{-|s_z|-}) \;.
\end{split}
\end{equation}
Therefore, Eq.~\eqref{eq_trbeta} still works but the factor $h/v$ has to be replaced by the following expression
\begin{equation}
\begin{split}
\frac{h}{v} \rightarrow \frac{\chi}{3}& \Bigl( 
\frac{\bracket{+|s_z|+} -\bracket{-|s_z|-}}{v_z}
+\frac{\bracket{+|s_x|-} + \text{c.c.}}{v_x} \\
&+\frac{i\bracket{+|s_y|-} + \text{c.c.}}{v_y}\Bigr)\,.
\end{split}
\end{equation}

It is also possible that the Weyl node is tilted. In this case, the Hamiltonian can be written as $\hat{H}=v^\prime k_x+\hat{H}_a$. Since neither the wave function nor the energy difference changes with the tilting, the response function in Eq.~\eqref{eq_betaz} is unaffected, so is the topological IFE. However, the topological IFE will be compromised in a type-II Weyl node with $|v^\prime|>\min\{|v_x|,|v_y|,|v_z|\}$, due to the presence of an open Fermi surface.

So far we have only considered the spin contribution to the magnetization.  In general, the orbital motion of electrons also contributes to the magnetization. For localized atomic orbitals, one should replace the spin operator $2\bm s$ by $\bm L+2\bm s$ in Eq.~\eqref{eq_betaf}, where the factor 2 is from the spin gyrotropic factor~\cite{Berritta2016}.  The resulting IFE is still topological and frequency-independent.  In addition, there are also contributions from the intercell orbital motion of electrons~\cite{Thonhauser2005, Xiao2010}.  It has been shown that at zero Fermi energy, the intercell orbital motion does not contribute to the IFE~\cite{Tokman2020}.

Finally, if there are several pairs of Weyl nodes, the resulting frequency-independent magnetization is simply the summation of that for each pair of Weyl nodes. Therefore, if the spin texture for different pairs of Weyl nodes are the same, the signal is amplified by the number of pairs. This is also a possible direction to manipulate the photoinduced magnetization.

\textit{Lattice model.}---In the following, we consider a lattice model of the Weyl semimetal to evaluate the induced magnetization from chiral lights and to demonstrate that the frequency-independence of the induced magnetization is robust against lattice effect. 

\begin{figure}[t]
  \includegraphics[width=\columnwidth]{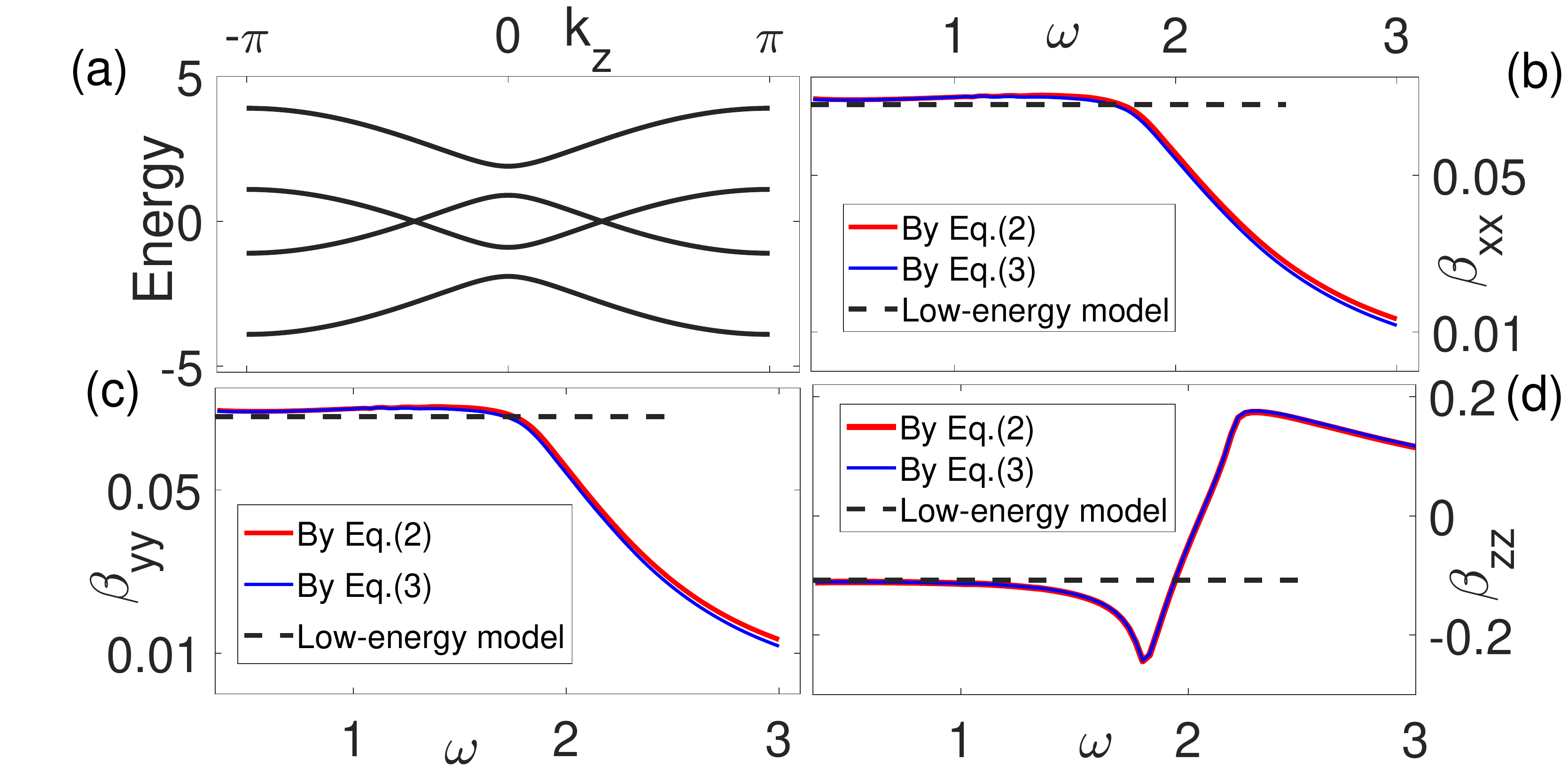}
  \caption{The inverse Faraday effect in a lattice model. Panel (a) is the spectrum showing two Weyl nodes. Panel (b)-(d) is the for $\beta_{xx}$, $\beta_{yy}$, and $\beta_{zz}$ calculated using Eq.~\eqref{eq_betaf} and Eq.~\eqref{eq_betaz} in the lattice model, and Eq.~\eqref{eq_betaz} in the low-energy effective model. Energy in (a) and frequency in (b-d) are in units of the model parameter $r$. $\beta$ is in units of $e^2\tau_0 \mu_B/(\hbar r a)$.}\label{fig_fig1}
\end{figure}

We consider a Weyl semimetal regularized on a cubic lattice, with the Hamiltonian given by~\cite{Koshino2016}
\begin{equation}\label{eq_latham}
\hat{H} = \lambda \tau_x (s_x \sin k_x+s_y \sin k_y +s_z \sin k_z)+\epsilon\tau_z+b s_z \;,
\end{equation}
where the Pauli matrices $\bm \tau$ and $\bm s$ operate in the orbital and spin space, respectively, $\epsilon=m+r(3-\cos k_x-\cos k_y-\cos k_z)$, $b$ is the Zeeman field, $\lambda$ is the strength of the spin-orbital coupling, and $\bm k$ is the dimensionless momentum~(i.e., we set the lattice constant $a=1$).  Equation~\eqref{eq_latham} describes a Dirac semimetal when $m=b=0$. In other cases, the Dirac node may split and Weyl nodes are possible. We set $\lambda=r$, $b=1.4 r$, and $m=0.5 r$.  There are two Weyl nodes located at $(0,0,\pm k_0)$ with $k_0$ satisfying $\cos k_0=(2+2m+m^2-b^2)/[2(m+1)]$, as shown in Fig.~\ref{fig_fig1}(a). 

Near the Weyl node at $(0,0, k_0)$, we extract the following effective Hamiltonian $\hat{H}=v_x k_x\sigma_x-v_y k_y\sigma_y+ v_z k_z \sigma_z\,,$
where $v_x=v_y=r$, $v_z=0.978 r$, and $\bm \sigma$ is in some pseudospin space. The basis for $\bm \sigma$ reads
\begin{equation}
\begin{split}
\ket{+} &= 0.343 \ket{A \uparrow} 
- 0.939 \ket{B \uparrow} \;, \\
\ket{-} &= -0.939 \ket{A \downarrow}
+ 0.343 \ket{B \downarrow} \;.
\end{split}
\end{equation}
where $A$ and $B$ are the orbital indices, and $\uparrow$ and $\downarrow$ are the spin indices. Since the basis possesses a spin structure, this Weyl node has a nonzero helicity based on Eq.~\eqref{eq_fhel}. Using Eq.~\eqref{eq_totbeta}, we obtain $\beta_{xx}=\beta_{yy}=0.068 \tau_0/r$ and $\beta_{zz}=-0.108\tau_0/r$. We see that $\beta_{xx}$ and $\beta_{zz}$ are quite different, although the corresponding Fermi velocities are almost the same. This difference is mainly due to different spin structures along the $x$ and $z$ directions.

We also calculate the response function using the full expression in Eq.~\eqref{eq_betaf} and the approximated one in Eq.~\eqref{eq_betaz} for the lattice model in Eq.~\eqref{eq_latham}, as shown in Fig.~\ref{fig_fig1} (b)-(d). We find that in the frequency-independent regime they agree well with the above calculation using the effective low-energy Hamiltonian. This confirms the validity of only considering the dominant contribution from the $n = m$ terms in Eq.~\eqref{eq_betaz}. It also suggests that the lattice effect generally affects the induced magnetization only slightly. When the light frequency is close to $2 r$, the optical transition occurs in the region where the two Weyl nodes start to intercept each other, and the frequency-independence is gradually destroyed by the lattice effect.

\textit{Ultrafast switching.}--- 
The photoinduced magnetization in Eq.~\eqref{eq_pm} is at saturation beyond the time scale of $\tau_0$, which is usually on the order of picosecond. However, to realize ultrafast control of the magnetization, it is often desirable to apply a strong but very short laser pulse on the time scale of sub-picosecond. In this ultrafast regime, the induced magnetization is far from saturation, and can be described by the following phenomenological equation
\begin{align}\label{eq_dyn}
\frac{dM_i}{dt}=\frac{\beta_{ij}}{\tau_0} (i\bm E\times \bm E^\star)_j-\frac{M_i}{\tau_0}\,,
\end{align}
In Eq.~\eqref{eq_dyn}, the first term on the right hand side is obtained in a similar way as the current injection in the circular photogalvanic effect~\cite{Sipe2000,Juan2017}. It represents the injection of magnetization by absorbing the circularly polarized light. The second term is a phenomenological collision term and $\tau_0$ thus has the meaning of the spin relaxation time. For illustration purpose, we consider a light field with a constant power. One can find that in the ultrafast regime~($t\ll\tau_0$), magnetization increases linearly, similar to the circular photogalvanic effect~\cite{Juan2017}. It then saturates to the magnetization given in Eq.~\eqref{eq_pm}.

We now estimate the typical time required to flip the inherent magnetization in a magnetic Weyl semimetal. We assume a laser fluence 1 mJ/cm$^2$ with a duration of 50~fs, a lattice constant $2$~\AA, and a hopping strength 1~eV. According to the calculation in the lattice model, we estimate that the induced magnetization is around $0.23 \mu_B$ per unit cell~\cite{suppl}. This induced magnetization affects the inherent magnetization through the following exchange coupling
\begin{align}
\hat{H}=J\bm S_{loc}\cdot \bm s_{el}\,,
\end{align}
where $J$ is the exchange coupling strength, $\bm S_{loc}$ is the local angular momentum, and $\bm s_{el}$ is the electronic angular momentum. 
Based on this coupling, the induced magnetization amounts to an effective field $\bm B_{eff}=J \bm m_{el}/ (2\mu_B)$ for the switching of $\bm S_{loc}$ through a field-like torque. 
Assuming a typical exchange coupling strength of $150~{\rm meV}$~\cite{Hsu2014}, we find that the time required to switch the direction of $\bm S_{loc}$ is $2\pi \mu_B \hbar/(J |\bm m_{el}|)=120$~fs~\cite{suppl}. Therefore, the topological IFE in Weyl semimetals is strong enough for ultrafast magnetization switching.

In summary, we show that in Weyl semimetals, the momentum-space helical spin texture can generate a frequency-independent inverse Faraday effect protected by topology. It is robust against both symmetry and lattice effect. It can be also used to realize ultrafast magnetization switching.  Our work shows that the helicity of the Weyl node can be very important, and calls for future study to further elucidate its implications.

\begin{acknowledgments}
We acknowledge insightful discussions with Dazhi Hou. Y.G. acknowledges support from the Startup Foundation from USTC.
Work at Carnegie Mellon is supported by the Department of Energy, Basic Energy Sciences, Grant No.~DE-SC0012509.  
\end{acknowledgments}

\bibliography{weylfar}

\end{document}